\documentclass[useAMS,usenatbib]{mn2e}
\usepackage{graphicx}
\usepackage{graphics}
\usepackage{epstopdf} 
\usepackage{float}
\usepackage{bm}
\usepackage{epsf}
\usepackage{url}

\title[Star Formation Isochrone Surfaces]{Star Formation Isochrone Surfaces: Clues on Star Formation Quenching in Dense Environments}
\author[Aragon-Calvo M.A. et al.]{M.A. Aragon-Calvo$^{1,2}$\thanks{E-mail:miguel@pha.jhu.edu}, Mark C. Neyrinck,$^{2}$, Joseph Silk$^{2,3}$\\
$^{1}$Department of Physics and Astronomy. University of California, Riverside, CA, USA.\\
$^{2}$ Department of Physics and Astronomy. Johns Hopkins University. Baltimore, MD 21218, USA.\\
$^{2}$Institut d Astrophysique de Paris. Univ. Paris VI, 98 bis boulevard Arago, 75014 Paris, France}
\begin{document}

\pagerange{\pageref{firstpage}--\pageref{lastpage}} \pubyear{2002}
\maketitle
\label{firstpage}

\begin{abstract}

The star formation history of galaxies is a complex process usually considered to be stochastic in nature, for which we can only give average descriptions such as the color-density relation. In this work we follow star-forming gas particles in a hydrodynamical N-body simulation back in time in order to study their initial spatial configuration. By keeping record of the time when a gas particle started forming stars we can produce \textit{gas-star isochrone surfaces} delineating the surfaces of accreting gas that begin producing stars at different times. These accretion surfaces are \textit{closely packed} inside dense regions, intersecting each other, and as a result galaxies inside proto-clusters stop accreting gas early, naturally explaining the color dependence on  density. The process described here has a purely gravitational / geometrical origin, arguably operating at a more fundamental level than complex processes such as AGN and supernovae, and providing a conceptual origin for the color-density relation.
\end{abstract}
\begin{keywords}
Cosmology: large-scale structure of Universe; galaxies: kinematics and dynamics, Local Group; methods: data analysis, N-body simulations
\end{keywords}

\section{Introduction}

The observed properties of galaxies are the combined result of complex internal mechanisms (secular evolution) such as supernovae, AGN feedback, etc. \citep{Powell11, Larson80}, and ii) external environmental mechanisms such as galaxy interactions and mergers, harassment, etc. \citep{Gunn72,Moore96,Kawata08}. The role of cosmic environment on star formation is evident in processes such as the morphology-density relation \citep{Dressler80} and the related color-density relation, which encode the effect of environment (density) on star formation history (color) \citep{Blanton05, Bell04,Baldry04,Thomas05,Dekel09}. While environmental processes leave their imprint on all galaxies, they are more clearly seen in dense environments such as massive galaxy clusters where star formation is mostly quenched. On the other hand, low-density environments contain the majority of star-forming galaxies.  Several mechanisms are assumed to contribute to the observed bimodality in the color distribution and the decreasing fraction of blue galaxies with increasing density,  such as galaxy mergers and harassment. \citep{Gunn72,Larson80,Moore96}. These mechanisms however, do not offer a direct link between star formation and environment. Even galaxy mergers, which seemed to explain the relation via the apparently related morphology-density relation, have been shown to play a minor role in color evolution \citep{Blanton05,Skibba09}. 

Galaxies accrete cold gas via narrow filamentary streams that penetrate deep into the galaxy \citep{Keres05,Dekel06,Dekel09,Voort11b}, fueling star formation shortly after accretion \citep{Bauermeister10}. The gravitational collapse of matter into the galaxy sets a natural order in the accretion of gas, i.e. nearby gas is accreted first while gas in distant reservoirs is accreted later. If star formation closely follows gas accretion, we should then expect a simple relation between star formation time and the original distance between the gas cloud that formed the stars and the galaxy, at least approximately. This may provide a link between the stellar populations of galaxies, encoded in their color, and the initial spatial configuration of the proto-galaxy. In this Letter, we explore this idea by tracking star-forming gas particles back to their initial position in two extreme cases, one isolated ``Milky Way" galaxy and one central galaxy inside a large cluster.

\begin{figure*}
  \centering
  \includegraphics[width=0.9\textwidth,angle=0.0]{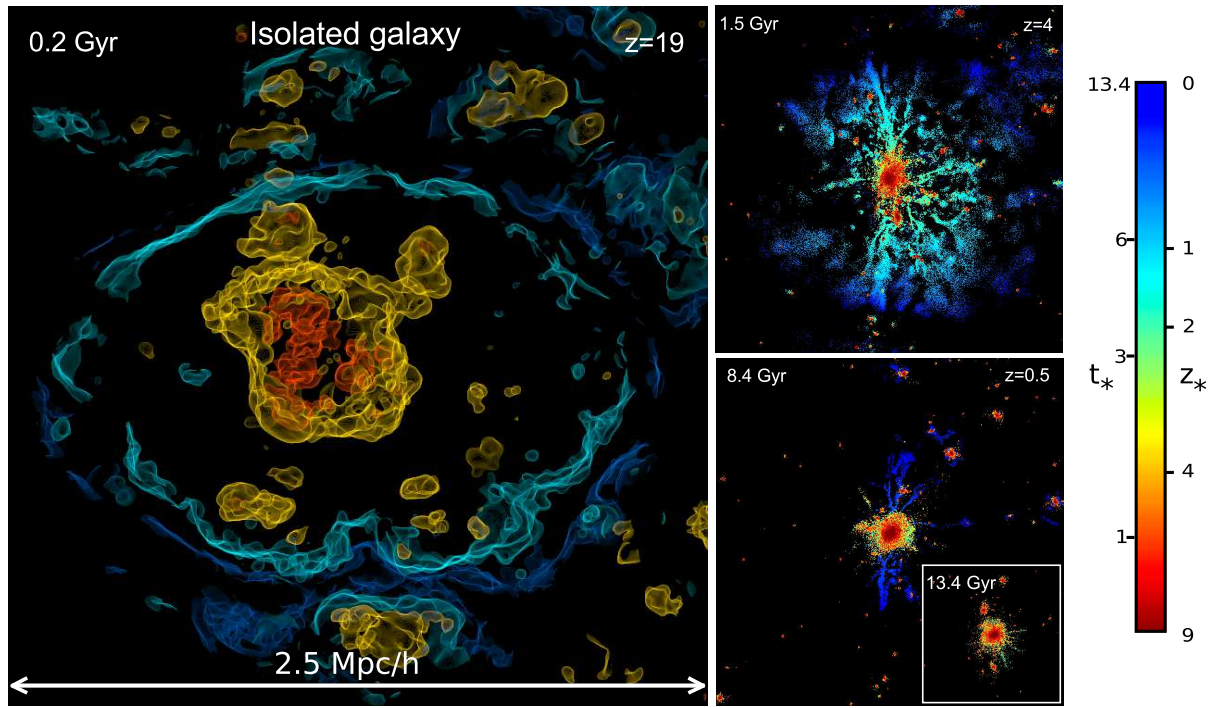}
  \caption{Gas$\to$star conversion times ($t_*, z_*)$ in an isolated ``Milky Way"  (MW) galaxy with a mass of $\sim 2 \times 10^{12} h^{-1} $M$_{\odot}$. We only show star-forming gas particles. Colors indicate the gas$\to$star conversion time using a color table that intuitively mimics the observed present-time stellar population colors. Left panel: four gas-star conversion isochrone surfaces at early times ($z=19$), corresponding to (from dark-blue to red) $t_* = 10.2,5.7,2.2$ and $1.2$ Gy. The MW galaxy, being relatively isolated, can freely accrete gas as shown in its spherical (in 3D) isochrones extending to near the present time (dark blue). The right panels show the subsequent evolution of the star-forming gas particles. The shell-based visualization was inspired by the work of \citet{Kahler02}.}
  \label{fig:000084}
\end{figure*} 

\section{Simulations}

The results presented in this work are based on two full hydrodynamic zoom resimulations: i) an isolated ``Milky Way" galaxy with a present-time mass of $2 \times 10^{12} h^{-1}$M$_{\odot}$ located inside a cosmological wall identified with the MMF-2 method \citep{Aragon14} from a 32 $h^{-1}$Mpc box and ii) a ``galaxy cluster" with a present-time mass of $3 \times 10^{14} h^{-1}$M$_{\odot (z=0)}$. selected from a 64 $h^{-1}$Mpc box. Both haloes were resimulated at high resolution by first selecting particles inside a sphere of 3 $h^{-1}$Mpc radius centered on the target halo identified at $z=0$. From the selected particles' initial positions we created a binary mask which was then filled with high-resolution particles and the rest of the box with layers of decreasing resolution particles. The mass resolution (corresponding to $1024^3$ particles) for the ``Milky Way" and  ``galaxy cluster" are $1.7\times10^{7} h^{-1}$M$_{\odot}$ and $2.2\times10^{6} h^{-1}$M$_{\odot}$  respectively. The Gadget-3 code used to run the resimulations implements simple recipes for hydrodynamics and chemical enrichment including stochastic star formation (for gas with a hydrogen number density  $> n_h = 0.1$ cm$^{-3}$), SN feedback and winds \citep{Springel03b}. Additionally we ran a simulation on a 32 $h^{-1}$ Mpc box including gas and star formation used to compute mean star formation time distributions (Fig. \ref{fig:isochrone_contours}). While more detailed star formation recipes are possible, in the present work their are not critical, as the main constraint on star formation we discuss here is geometric in nature. 

\begin{figure*}
  \centering
  \includegraphics[width=0.9\textwidth,angle=0.0]{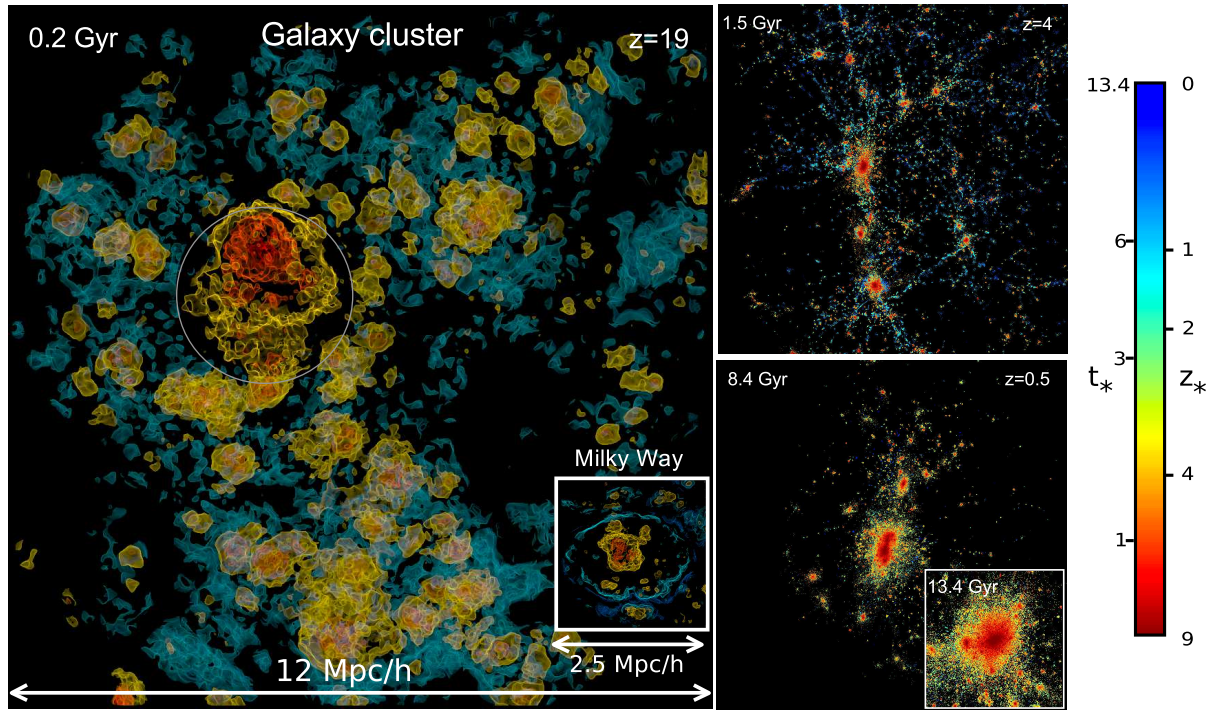}
  \caption{Gas-star conversion times in a galaxy cluster with mass of $\sim 3 \times 10^{14} h^{-1} $M$_{\odot}$.  (see Fig. \ref{fig:000084}). Star-forming gas is accreted from increasingly distant isochrone surfaces, centered in the progenitors of massive galaxies (red blobs). Proto-cluster galaxies, being closely packed, are surrounded by a layer of galaxies and are effectively isolated from late star-forming gas (light blue isochrone surfaces). The white circle in the top panel shows the central proto-galaxy in a window carved through the gas cloud. For comparison,  we show the Milky Way galaxy from Fig. \ref{fig:000084} (small white square) on the left panel.}
  \label{fig:000000}
\end{figure*}

\section {Tracing back stellar populations in time via isochrone surfaces}

In order to investigate the connection between primordial environment, gas accretion and star formation history, we identified and followed star-forming gas particles, i.e. gas particles that at some point during the simulation's history produced stars, from the present time back to the initial conditions. The question we are trying to answer is: where did the gas that produced different star populations inside a given galaxy come from? For each gas particle we also stored the time when it started producing stars, here referred to as the gas$\to$star conversion time, $t_*$. When convenient, we will use the equivalent gas$\to$star conversion redshift $z_*$ and scale factor $a_*$. In cases where several star particles  were spawned from the progenitor gas particle or when the gas particle was completely converted to star particles we assigned $t_*$ to the first time the gas particle produced a star. By doing so, we were able not only to follow stars after they are formed, but to trace their ``progenitor" gas particles back to their initial comoving positions (Lagrangian coordinates) and study their initial spatial arrangement.

We converted the discrete gas particle positions to a continuous scalar field containing the interpolated values of $t_*$ at each gas particle's position. We used the Delaunay-based interpolation scheme described in \citet{Bernardeau96} in which a scalar value ($t_*$) defined at each sampling point (gas particles) is linearly interpolated inside the tetrahedra defined by the point distribution. The tessellation was computed from the particle positions at $z=19$ instead of the true Lagrangian positions in order to alleviate the degenerate cases arising from computing the Delaunay tessellation on a regular grid of particles. The $t_*$ field was interpolated into a regular grid. From the $t_*$-field we computed iso-surfaces at a set of gas$\to$star conversion times. These \textit{gas$\to$star isochrone surfaces}, $\mathcal{S}_{t_*}$  define regions of gas that, after being accreted into galaxies, started forming stars at the same time.  

\section{Results}

We begin discussing the case of the relatively simple ``Milky Way" galaxy and then proceed to the more complex galaxy cluster. Figure \ref{fig:000084} shows the star-forming evolution of a galaxy in isolation. The left panel shows several gas$\to$star isochrone surfaces color-coded with time. This technique allows us to see the full time evolution and spatial arrangement of star-forming gas particles in one single frame. The earliest star forming gas is the closest to the proto-galaxy and the first to be accreted (red surfaces). Subsequent layers of gas are accreted and form stars at later times. The simple accretion history of this galaxy is reflected in its also simple isochrone surfaces. There is one single large proto-galaxy and radial semi-spherical shell extending in Lagrangian space. Since there is no impediment for this galaxy to accrete star-forming gas, it can continue forming stars until the present time as indicated by the existence of the outer isochrone surface corresponding to $z=0$ (dark blue). The subsequent evolution of the particle distribution is shown in the right panels. Note that while the isochrone surfaces are semi-spherical shells in Lagrangian space they correspond to filaments in configuration (Eulerian) space. This technique allows us to directly see that star-forming gas is accreted mainly through narrow filamentary streams as first reported by \citet{Dekel09}.

Figure \ref{fig:000000} shows the \textit{star formation isochrone surfaces} for a galaxy cluster. The central proto-galaxy is the prominent structure near the center of the proto-cluster. For convenience in what follows, we will call this the central galaxy and the rest the satellite galaxies.  The isochrone surfaces are remarkably regular and there is a clear relation between gas-star conversion time and radial distance from centers of proto-galaxies even for this complex cluster (a triple major merger and several substructures, see right panels in Fig. \ref{fig:000000}). As in the case of the isolated galaxy, early star-forming gas is accreted first and so $\mathcal{S}_{z_*=9}$ (red surfaces) mark the centers of galaxies. The $\mathcal{S}_{z_*=9}$ surfaces enclose a volume of gas roughly 8 times larger than the same isochrone for the Milky Way galaxy, indicating a star formation rate 8 times larger for the central galaxy compared to the Milky Way. 
The $\mathcal{S}_{z_*=3}$ surfaces (yellow) surrounding satellite galaxies are relatively isolated compared to the $\mathcal{S}_{z_*=3}$ surfaces around the central galaxy where surfaces from adjacent galaxies are intersecting. The central galaxy, being surrounded by a compact shell of adjacent satellite galaxies is geometrically constrained to accrete star-forming gas beyond the $\mathcal{S}_{z_*=3}$ surface. This can be seen in the almost total lack of  $\mathcal{S}_{z_*=1}$ surfaces (light blue) around the central galaxy. Satellite galaxies on the other hand are still able to accrete gas at this time although there is no noticeable accretion of star-forming gas after $z\sim1$. 

From Figures  \ref{fig:000000} and  \ref{fig:000084} we can begin to see a simple model: as a galaxy grows, it carves out increasingly large surfaces (in Lagrangian coordinates) of star­-forming gas from it surroundings. These surfaces extend from the center of the proto-galaxy until the maximum radius of influence from which a galaxy can gravitationally accrete mass. Galaxies in dense environments can be seen as a \textit{closely packed system} of $\mathcal{S}_{z_*}$ spheres where adjacent galaxies compete for the available gas as they ``carve" the proto-cluster's volume. The intersection of isochrone surfaces from adjacent galaxies marks the time where no more gas is available for accretion and star formation. This occurs roughly at half the mean inter-galaxy separation, imposing a fundamental geometric limit to gas accretion in dense environments. On the other hand, galaxies in the outskirts of proto-clusters can have an extended star formation history due to their access to gas in the vicinity of the proto-cluster \citep{Papadopoulos01,Riechers10,Wolfe13} as observed in the different quenching times between central and satellite galaxies in groups\citep{Tal14}.

\begin{figure}
  \centering
  \includegraphics[width=0.4\textwidth,angle=0.0]{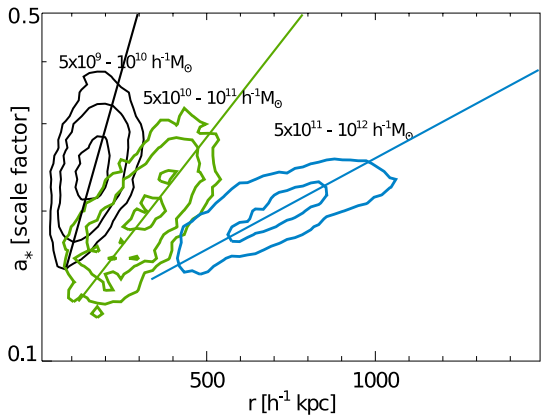}
  \caption{Distribution of Gas-star conversion times (in units of the scale factor) as function of initial distance from centers of proto-galaxies identified at $z = 4$ for three halo mass ranges. By extrapolating the intersection between the dotted lines (shown for comparison purposes) and the scale factor $a=1$ we can infer the maximum radius that a galaxy would carve at the present time under no geometric or other constraints.}
  \label{fig:isochrone_contours}
\end{figure} 

%
\subsection{Distribution of gas$\to$star conversion times}

Figure \ref{fig:isochrone_contours} shows the distribution of $t_*$ vs. initial comoving distance $r$ from the center of proto-galaxies. Since star formation closely follows gas accretion \citep{Bauermeister10} it should not be surprising that  $t_*$ is a monotonically increasing function of $r$. In the simplest case of a free­-falling gas particle, $t_*$ depends on the Lagrangian distance from the gas particle to the center of the proto-galaxy and the mass of the galaxy $M$, as $t_* \propto \sqrt{r^3 / M}$. In reality gas accretion is far more complex, involving both gravitational and hydrodynamical processes. We approximate the distribution in Fig. \ref{fig:isochrone_contours} by the empirical fit:

\begin {equation}\label{eq:fit}
\log_{10}(a_*) = -­0.4 ­-0.04 \log_{10} M + 3.6 r ­ - 0.3 r \log_{10} M,
\end{equation}

\noindent where $a_*$ is the time of gas­$\to$star conversion in units of the expansion factor, and $M$ is the mass of the galaxy, here identified at $z=4$, when the $r­ - a_*$ relation is still clearly defined. For galaxies identified at later times, the relation still holds, but non-linear interactions increase the dispersion. The fit provides an approximate model for star formation times that depends only on the mass and initial distance from the proto-galaxy's center, while ignoring feedback processes that further regulate star formation. In an isolated galaxy the accretion of star-forming gas is limited (in our simple model) only by the background cosmology and surrounding gravitational field. The extrapolated values of $r$ at $a_*=1$ are close to the radius containing the total mass of the halo in the mass range.

\begin{figure*}
  \centering
  \includegraphics[width=0.99\textwidth,angle=0.0]{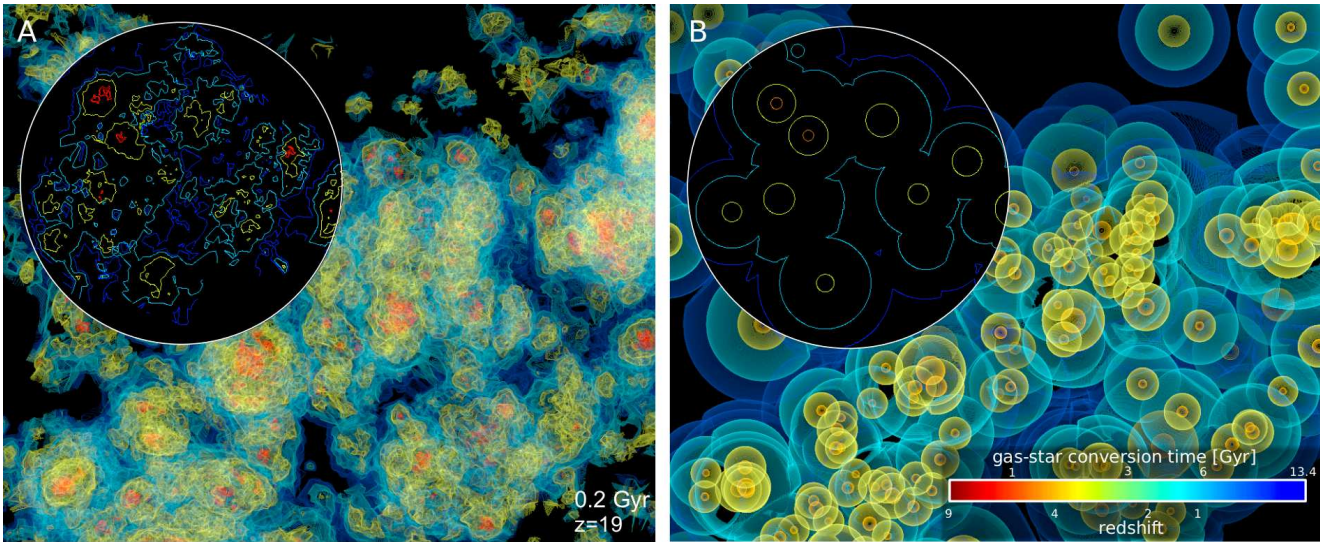}
  \caption{Gas-star conversion isochrones from the full N-body hydro simulation (left) and a Voronoi model (right) in a region of  $12 \; h^{-1}$ Mpc of side (horizontal axis). For clarity we show  two-dimensional slices through the isochrones inside the white circles at upper left.}
  \label{fig:voronoi_model}
\end{figure*} 

\section{Voronoi Modeling of Isochrone Surfaces}

The competition for gas by adjacent proto-galaxies is similar to other physical processes that result in a Voronoi segmentation \citep{Okabe00}. Using equation \ref{eq:fit}, we can model the isochrone surfaces as a system of \textit{random closely-packed spheres}. Assuming accretion through spherical Lagrangian shells centered at the proto-galaxy's position, we partition the space between proto-galaxies (identified at $z=4$ with masses $> 10^{10}  \; h^{-1}$M$_{\odot}$) using a weighted Voronoi tessellation in which the volume inside Voronoi cells define regions of gas accretion. The weight $w$ is given by the isochrone radius extrapolated to the present time (from equation \ref{eq:fit}):

\begin{equation}
	w = r_{a_*=1} = \frac{ 0.3702 + 0.0442 \; \log_{10}M } {3.577 - 0.2774 \; \log_{10}M }
\end{equation}

\noindent  We then generate a continuous $t_*$ time field by evaluating equation \ref{eq:fit} inside the Voronoi cell of each galaxy. Figure \ref{fig:voronoi_model} shows a side-by-side comparison between the measured isochrones and the ones obtained from the Voronoi model. The correspondence between the hydrodynamic simulation and the Voronoi model is remarkable. 
It may seem surprising that such a simple model based on geometric constraints agrees at all with the result of complex gravitational and hydrodynamical processes but in a sense it is also expected since at cluster-size scales, gas in the early universe basically follows dark matter. Early star formation has a strong deterministic component given by geometric constraints.

\section{Discusion and conclusions}

The most straightforward consequence of ``galaxy close packing"  is that the massive galaxies which are the progenitors of present-time groups and clusters, being surrounded by gas-competing galaxies, become cut off from their gas supply, becoming  ``quenched" already at early times \citep{Best97,Thomas05}. Massive cluster galaxies are located in environments where the number density of galaxies is several  times the average \citep{Daddi00}.  At $z=4$, the mean separation between $M > 10^{10} h^{-1}$M$_{\odot \;(z=4)}$ galaxies inside proto-­clusters is $d_{m} \simeq 0.6$ Mpc. From Fig. \ref{fig:isochrone_contours} we see that at $z=4$ haloes of mass between $10^{10-11} h­^{-1} $M$_{\odot (z=4)}$ and $10^{11-12} h­^{-1} $M$_{\odot (z=4)}$ accrete star-­forming gas from isochrone surfaces of $\sim0.5 h^{-1}­$Mpc and $\sim0.7 h­^{-1}$ Mpc radius respectively. As early as $z=4$, the radius of the corresponding isochrone surface for those galaxies is already of the same order as their mean separation, indicating the time of quenching for the central galaxy.

The mechanism described here offers a conceptual origin for the observed color-density relation by limiting gas accretion and star formation in dense environments. Galaxies in low-density regions, on the other hand, are not geometrically constrained and can, in principle, freely continue to accrete gas. However, the dominant role of the background cosmology, tidal fields from nearby structures and super-Hubble dynamics in low-density environments, can reduce and even halt gas accretion. In addition to this, other processes such as AGN and SN feedback play an important role.
Finally, the  model presented here may also shed some light on the observed cosmic star formation rate history. The peak in the $t_*$ distributions in Fig. \ref{fig:isochrone_contours} lie in the range $z=4-3$, close to the time when isochrone surfaces of massive galaxies (which produce most of the stars at that time) begin to intersect, as discussed above. It also corresponds to the observed peak in the cosmic star formation rate history \citep{Heavens04}. This peak, and subsequent drop in the star formation rate, can be interpreted as the result of the change in the main gas accretion mode from being gas accretion-driven (before isochrone intersection) to a less effective mode driven by galaxy mergers \citep{Voort11a,Feldmann14}.

\section{Acknowledgements}
This research was funded by a Big Data UC Riverside seed grant, the Betty and Gordon Moore foundation and a New Frontiers of Astronomy and Cosmology grant from the Templeton Foundation. Miguel Aragon would like to thank Bernard Jones for valuable comments and Volker Springel for the Gadget3 N-­body code.

\bibliography{references}
\bibliographystyle{hapj} 

\end{document}